\documentclass[aps,prb,reprint,superscriptaddress]{revtex4-2}

\usepackage{graphicx}

\begin{document}

\title{Specific heat of CeRhIn$_5$ in high magnetic fields: Magnetic phase diagram revisited}

\author{S.~Mishra}
\affiliation{Laboratoire National des Champs Magn\'{e}tiques Intenses (LNCMI-EMFL), CNRS, UGA, 38042 Grenoble, France}

\author{A.~Demuer}
\affiliation{Laboratoire National des Champs Magn\'{e}tiques Intenses (LNCMI-EMFL), CNRS, UGA, 38042 Grenoble, France}

\author{D.~Aoki}
\affiliation{Institute for Materials Research, Tohoku University, Oarai, Ibaraki, 311-1313, Japan}

\author{I.~Sheikin}
\email[]{ilya.sheikin@lncmi.cnrs.fr}
\affiliation{Laboratoire National des Champs Magn\'{e}tiques Intenses (LNCMI-EMFL), CNRS, UGA, 38042 Grenoble, France}

\date{\today}

\begin{abstract}
CeRhIn$_5$ is a prototypical antiferromagnetic heavy-fermion compound, whose behavior in a magnetic field is unique. A magnetic field applied in the basal plane of the tetragonal crystal structure induces two additional phase transitions. When the magnetic field is applied along, or close to, the $c$ axis, a new phase characterized by a pronounced in-plane electronic anisotropy emerges at $B^* \approx$ 30~T, well below the critical field, $B_c \simeq$ 50~T, to suppress the antiferromagnetic order. The exact origin of this new phase, originally suggested to be an electronic-nematic state, remains elusive. Here we report low-temperature specific heat measurements in CeRhIn$_5$ in high static magnetic fields up to 36~T applied along both the $a$ and $c$ axes. For fields applied along the $a$ axis, we confirmed the previously suggested phase diagram, and extended it to higher fields. This allowed us to observe a triple point at $\sim$ 30~T, where the first-order transition from an incommensurate to commensurate magnetic structure merges into the onset of the second-order antiferromagnetic transition. For fields applied along the $c$ axis, we observed a small but distinct anomaly at $B^*$, which we discuss in terms of a possible field-induced transition, probably weakly first-order. We further suggest that the transition corresponds to a change of magnetic structure. We revise magnetic phase diagrams of CeRhIn$_5$ for both principal orientations of the magnetic field based entirely on thermodynamic anomalies.
\end{abstract}

\maketitle

\section{INTRODUCTION}

Strongly correlated electron systems, such as high-temperature superconductors, iron-based superconductors, and heavy-fermion compounds, are of much experimental and theoretical interest. In all these materials, unconventional superconductivity is believed to emerge in the vicinity of a quantum critical point, a zero-temperature continuous phase transition. In addition, some of these materials host even more exotic phases. The latter include the pseudogap phase in high $T_c$ superconductors~\cite{Keimer2015}, an electronic-nematic state in iron-based superconductors~\cite{Fernandes2014}, and the mysterious ``hidden order" phase in the heavy-fermion compound URu$_2$Si$_2$~\cite{Mydosh2020}. These phases are still poorly understood, and their possible relation with unconventional superconductivity is a subject of much theoretical debate.

From the experimental point of view, Ce-based heavy-fermion materials are particularly suitable systems to investigate. In these compounds, the strength of the electronic correlations can be tuned by pressure, doping, and magnetic fields. Relatively small energy scales allow these materials to be tuned to quantum critical points by accessible pressures and magnetic fields. Unconventional superconductivity and/or other unusual states are often observed in these systems in the vicinity of a quantum critical point.

CeRhIn$_5$ is one of the best-studied heavy-fermion compounds. It crystallizes in the tetragonal HoCoGa$_5$ structure (space group $P4/mmm$), which can be viewed as a stack of alternating layers of CeIn$_{3}$ and RhIn$_{2}$ along the $c$ axis. The electronic specific heat coefficient, $\gamma \approx$ 400~mJ/K$^2$mol, makes CeRhIn$_5$ a moderate heavy-fermion material \cite{Hegger2000,Takeuchi2001,Kim2001}. At ambient pressure and zero magnetic field, it undergoes an antiferromagnetic (AFM) transition at $T_N =$ 3.8~K. Within the AFM phase, the Ce moments are antiferromagnetically aligned within the CeIn$_{3}$ planes. The moments spiral transversally along the $c$ axis with a propagation vector $\mathbf{Q} = (0.5, 0.5, 0.297)$ incommensurate with the crystal lattice~\cite{Bao2000}.

 A magnetic field applied in the basal plane of CeRhIn$_5$ induces two additional transitions, observed in specific heat~\cite{Cornelius2001}, thermal expansion, and  magnetostriction measurements~\cite{Correa2005}. The lower temperature transition is first order. It occurs at $B_{m}$ $\sim$ 2~T at low temperatures, and corresponds to a change of magnetic structure from incommensurate to commensurate~\cite{Raymond2007}. The higher temperature transition is second order. It corresponds to a change of the ordered moment, while the propagation vector, almost the same as in a zero magnetic field~\cite{Raymond2007}, becomes temperature-dependent~\cite{Fobes2018}. Both transitions were traced up to 18~T in static field measurements~\cite{Correa2005}. More recently, specific heat measurements in pulsed fields applied along the $a$ axis revealed a non-monotonic field dependence of $T_N$~\cite{Jiao2015}. The AFM transition temperature initially increases up to about 12~T, and then decreases monotonically all the way up to the critical field $B_{c} \sim 50$~T. In these measurements, however, the two field-induced phases were not observed. Therefore, the complete phase diagram for fields along the $a$ axis remains elusive.

When a magnetic field is applied along the $c$ axis, $T_N$ monotonically decreases until it is completely suppressed at $B_{c} \sim 50$~T~\cite{Jiao2015,Jiao2019}. Surprisingly, the critical field for this orientation is approximately the same as along the $a$ axis in spite of a considerable crystallographic and magnetic anisotropy of CeRhIn$_5$. On the other hand, the critical field was extrapolated from specific heat measurements in pulsed magnetic fields, while there is a difference between the results obtained in pulsed~\cite{Jiao2015,Jiao2019} and static~\cite{Kim2001} fields.

The most interesting feature was observed in various measurements at $B^* \simeq$ 30~T for a field applied either along or slightly tilted from the $c$ axis~\cite{Jiao2015,Moll2015,Ronning2017,Rosa2019,Lesseux2020,Kurihara2020,Helm2020}. While it was interpreted as a transition into an electronic-nematic state~\cite{Ronning2017}, the exact origin and nature of this anomaly is still under debate. Surprisingly, specific heat measurements have so far failed to show a direct indication of this anomaly~\cite{Kim2001,Jiao2015,Jiao2019}. It is thus still unclear whether the anomaly corresponds to a real thermodynamic phase transition or a crossover.

\begin{figure}[h!]
\includegraphics[width=\columnwidth]{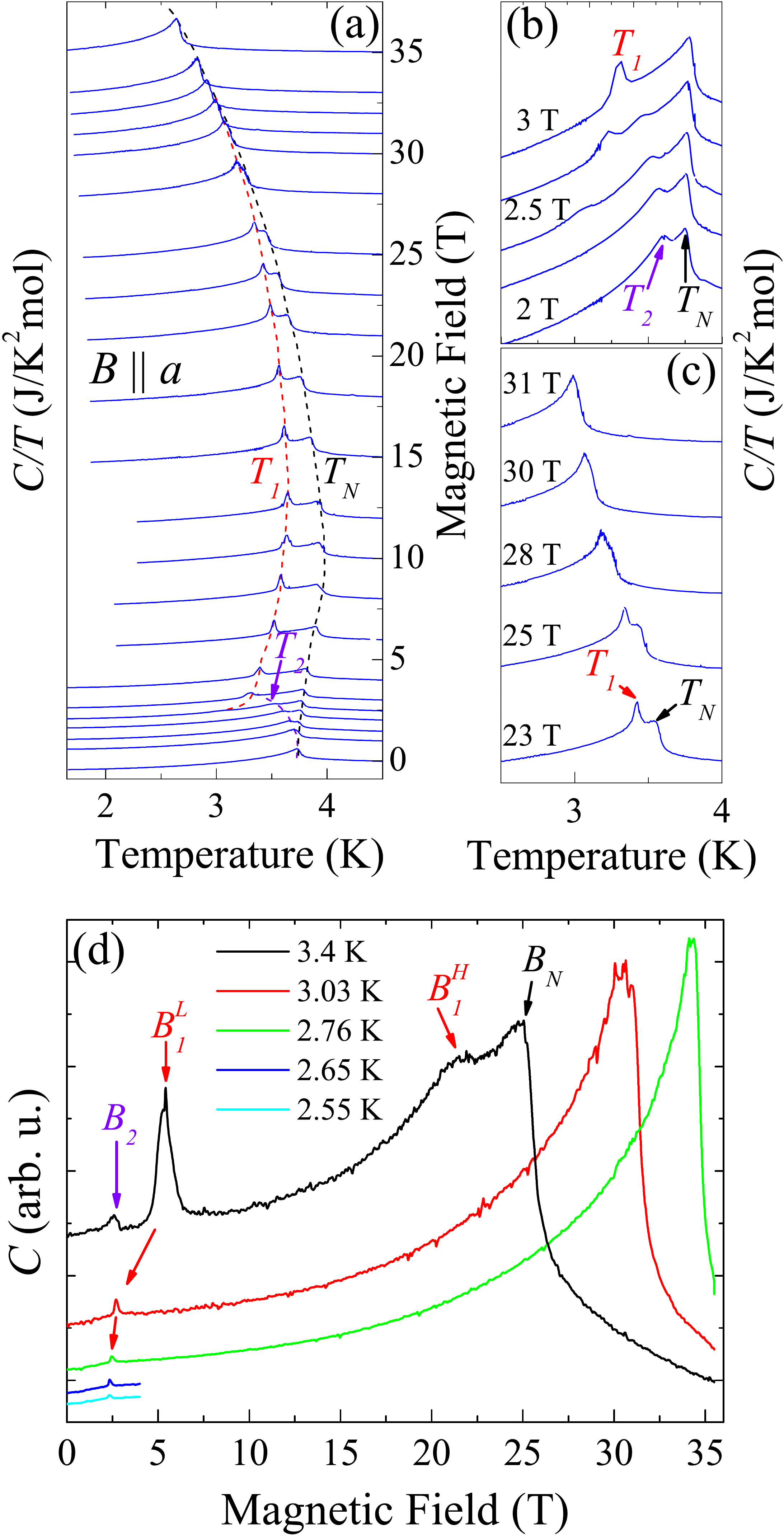}
\caption{\label{fig:Balonga} Specific heat divided by temperature, $C/T$,  of CeRhIn$_5$ for a magnetic field applied along the $a$ axis. (a) $C/T$ as a function of $T$ obtained from relaxation technique for several values of a magnetic field. Curves are vertically shifted according to the magnetic field scale shown in the right axis. A zoom at low and high fields is shown in (b) and (c), respectively. (d) Total specific heat, $C$, obtained from field sweeps using the AC technique.}
\end{figure}

The above mentioned inconsistencies and shortcomings demonstrate a clear need to perform specific heat measurements in high static fields to ascertain the quantum critical point, complete the phase diagram for fields along the $a$ axis, verify the phase diagram along the $c$ axis, and seek a direct  evidence for the enigmatic novel state at $B^*$.

In this paper, we report high-field low-temperature specific heat measurements on a single crystal of CeRhIn$_{5}$. The measurements were performed in static fields up to 36~T for field orientations both along the $a$ and $c$ axes. For a field applied along the $a$ axis, we observed all the previously reported transitions, and traced them to higher fields. For a field along the $c$ axis, we observed the so far elusive anomaly at $B^*$ in addition to the AFM transition. Based on these features observed in specific heat, we propose a revision of the magnetic phase diagram of CeRhIn$_{5}$ for both principal orientations of the magnetic field.

\section{EXPERIMENTAL DETAILS}

The high-quality single crystal of CeRhIn$_{5}$ with the dimensions of 1.3$\times$0.8$\times$0.2~mm$^3$ (the length of the sample is parallel to the $c$ axis) and a mass of 1.55~mg  used in the present study was grown by the In-self-flux technique, details of which can be found elsewhere \cite{Shishido2002}. Specific heat measurements were performed in static magnetic fields to 36~T by either a thermal relaxation technique at constant field or AC technique at constant temperature~\cite{Suppl}.

\nocite{Lortz2007,Sullivan1968,Baloga1977}

\section{RESULTS AND DISCUSSION}

\begin{figure}[htb]
\includegraphics[width=\columnwidth]{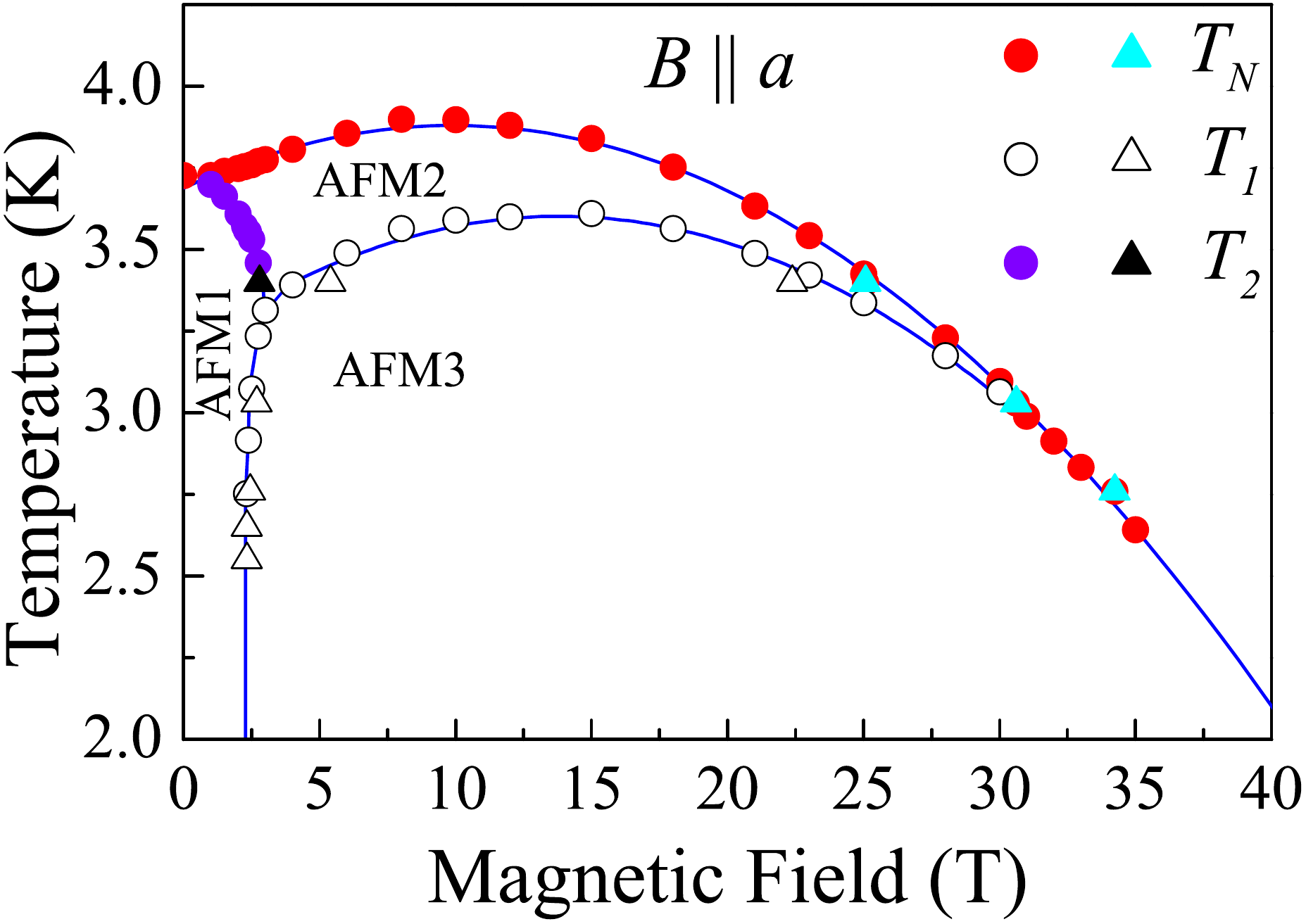}
\caption{\label{fig:PDa} Magnetic phase diagram of CeRhIn$_5$ obtained from relaxation (circles) and AC (triangles) specific heat measurements for a field applied along the $a$ axis. Closed and open symbols correspond to second- and first-order transitions, respectively.}
\end{figure}

Figure~\ref{fig:Balonga}(a) shows specific heat divided by temperature, $C/T$, obtained from relaxation measurements for a magnetic field applied along the $a$ axis. For this field orientation, apart from the AFM transition at $T_N$, there are two additional field-induced transitions at $T_1$ and $T_2$, as shown in Fig.~\ref{fig:Balonga}(b). The transition at $T_1$ manifests itself by a sharp $\delta$-like peak characteristic of a first-order transition. The transition at $T_{2}$ appears as a $\lambda$-type anomaly typical for a second-order transition. This transition is observed only at low fields, as shown in Fig.~\ref{fig:Balonga}(b). In agreement with previous reports, $T_N$ initially increases up to about 10~T, and then decreases monotonically up to the highest field of our measurements. The transition temperature $T_1$ shows a similar trend. Above 3~T, $T_1$ increases up to about 12~T, and then starts to decrease. Its suppression rate, however, is slower than that of $T_N$. With increasing field, the two transitions approach each other. At 28~T, the two transitions are barely distinguishable, and at 30~T only the transition at $T_N$ remains, as shown in Fig.~\ref{fig:Balonga}(c).

All the transitions are also observed in measurements using the AC technique, as shown in Fig.~\ref{fig:Balonga}(d). In particular, the curve obtained at 3.4~K, shows four phase transitions. Interestingly, while the low-field transition at $B_1^{L}$ manifests itself as a sharp peak, its high-field counterpart at $B_1^{H}$ appears as a rather smeared anomaly, although it is also a first-order transition. This is not surprising considering that standard AC calorimetry models are based on steady-state measurements, which is not the case for a first-order phase transition due to the involvement of latent heat. Therefore, a first-order transition does not always manifest itself as the canonical $\delta$-like feature. The shape, and even the presence, of an anomaly depends on the latent heat associated with the transition. On the other hand, a second-order transition always manifest itself as a distinct $\lambda$-like anomaly, which is, indeed, the case for the two known second-order transitions at $B_2$ and $B_N$.

The resulting magnetic field temperature, $B-T$, phase diagram for field along the $a$ axis is shown in Fig.~\ref{fig:PDa}. It contains three different antiferromagnetic phases labeled AFM1, AFM2, and AFM3. The magnetic structure of all three phases was previously determined by neutron diffraction~\cite{Raymond2007,Fobes2018}. The zero-field phase AFM1 corresponds to an incommensurate antiferromagnetic spin helix with a propagation vector $\mathbf{Q} = (0.5, 0.5, 0.297)$. The AFM2 phase is an incommensurate elliptical helix with strongly modulated magnetic moments and a temperature-dependent propagation vector. The AFM3 phase is a commensurate collinear square wave (`up-up-down-down' configuration) with a propagation vector $\mathbf{Q} = (1/2,1/2,1/4)$. All three phases meet at a triple point inside the AFM phase at (3~T, 3.4~K). The AFM2 phase exists only in a narrow temperature range close to $T_N$. This range shrinks with increasing magnetic field until the AFM2 phase is completely suppressed at $\sim$ 30~T, giving rise to yet another triple point. Remarkably, this field is about the same as $B^*$, at which the putative electronic-nematic phase emerges for fields close to the $c$ axis. Above 30~T, only the commensurate phase AFM3 exists up to the complete suppression of the AFM order. A naive quadratic fit of $T_{N}$ vs $B$ reveals a critical field $B_{c} \simeq$ 54~T, in agreement with previous pulsed field results~\cite{Jiao2015}.

\begin{figure}[htb]
\includegraphics[width=\columnwidth]{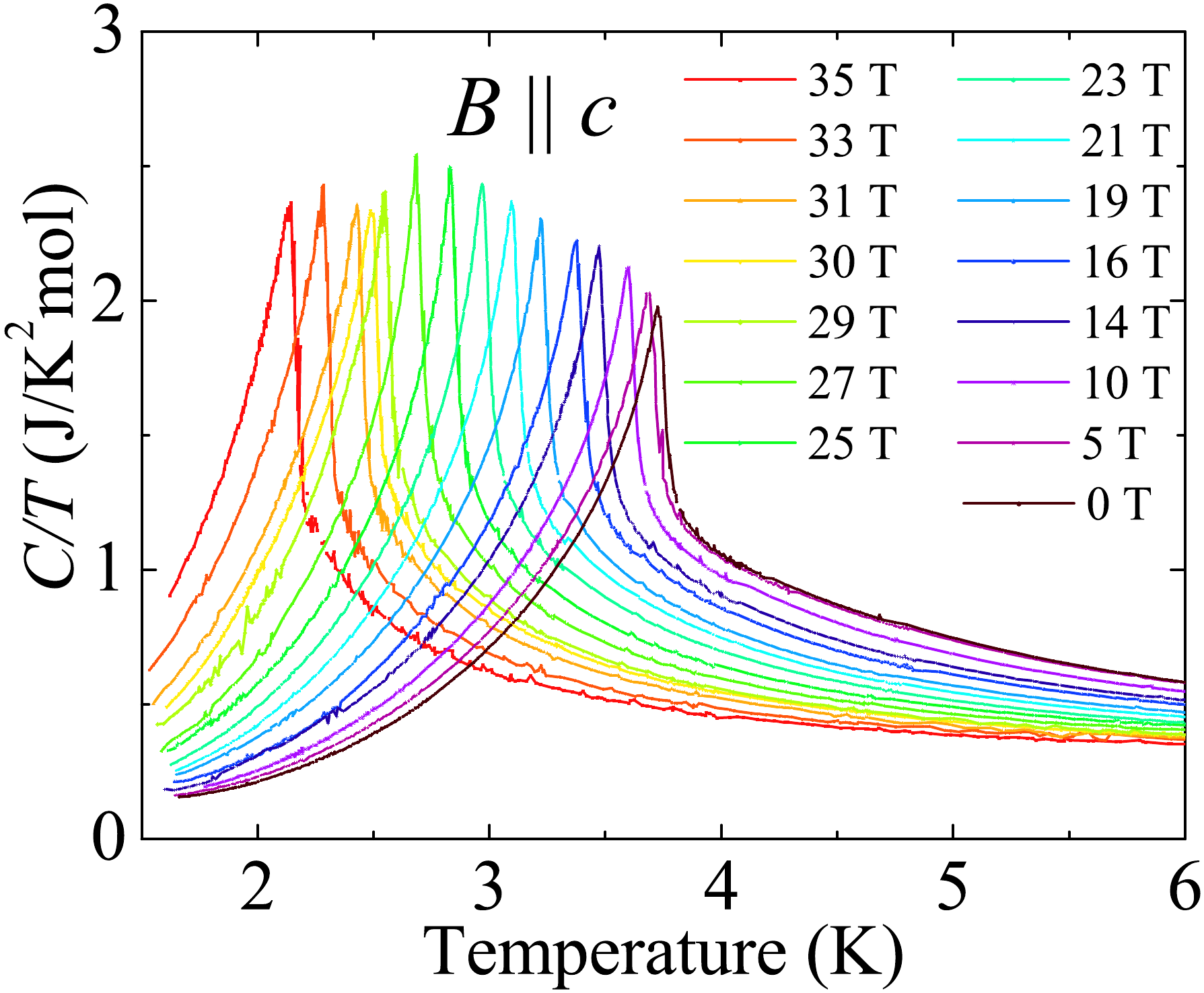}
\caption{\label{fig:Balongc} $C/T$ of CeRhIn$_5$ for a magnetic field applied along the $c$ axis obtained from temperature sweeps using the relaxation technique.}
\end{figure}

Figure~\ref{fig:Balongc} shows the temperature dependence of the specific heat divided by temperature, $C/T$, obtained using the relaxation technique at different magnetic fields applied along the $c$ axis. For this orientation of the magnetic field, $T_{N}$ is gradually suppressed, consistent with previous reports~\cite{Kim2001,Jiao2015,Jiao2019}. This is the usual behavior observed in AFM heavy-fermion compounds. However, we observed a nonmonotonic behavior of the specific heat jump at the AFM transition. With increasing field, the jump size gradually increases up to 27~T, above which there is a small abrupt drop. The jump size then remains almost constant between 29 and 35~T, the highest field of our measurements. A similar behavior was also observed in the previous studies~\cite{Jiao2019,Cornelius2001}. This unusual behavior indicates that there might be a change in the AFM state between 27 and 29~T.

\begin{figure}[htb]
\includegraphics[width=\columnwidth]{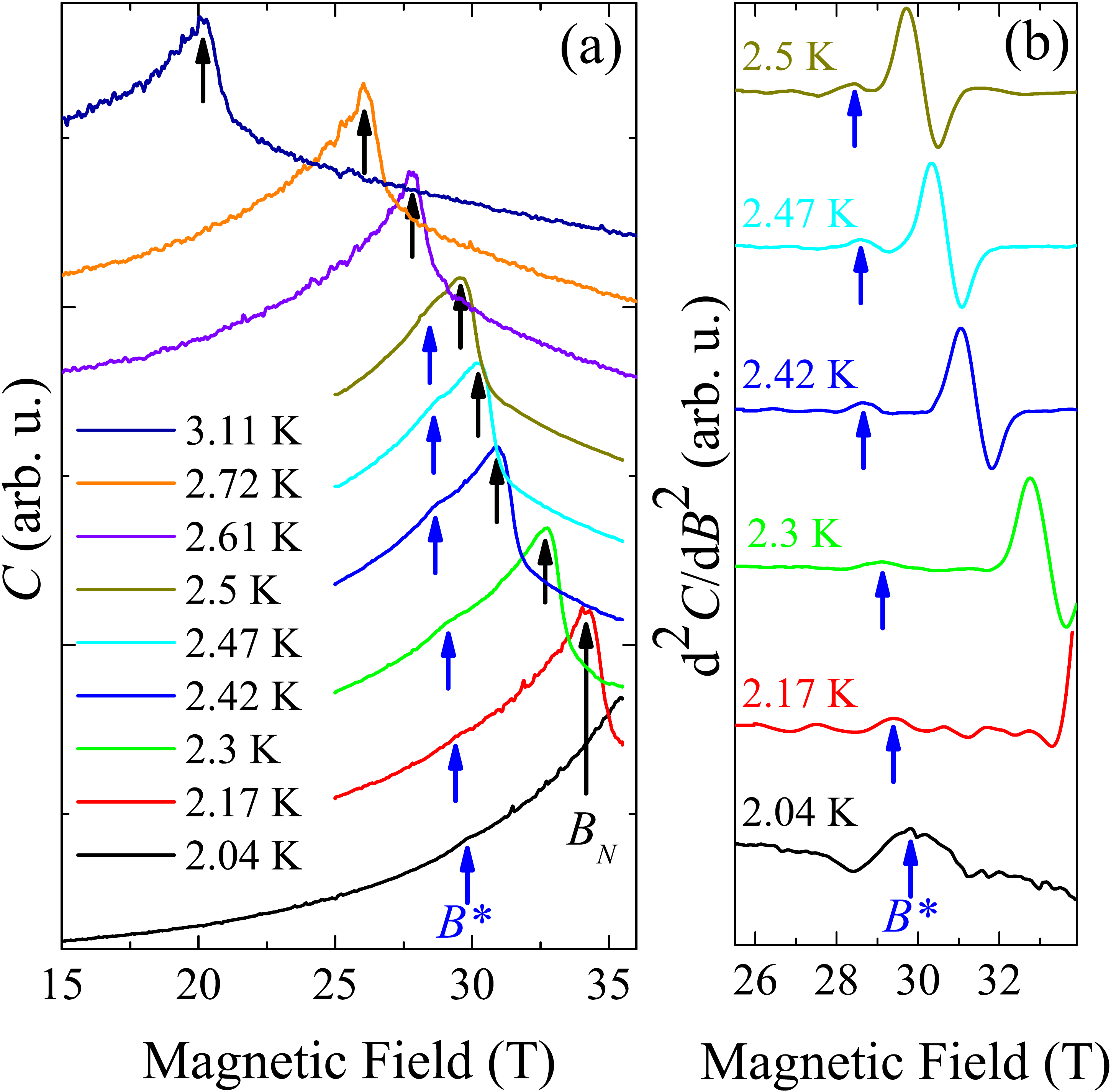}
\caption{\label{fig:BalongcAC} (a) Specific heat of CeRhIn$_5$ for a magnetic field applied along the $c$ axis obtained from field sweeps using the AC technique. Curves are vertically shifted for clarity. (b) Second derivatives of the heat capacity shown in (a) with respect to a magnetic field. Arrows indicate the AFM transition and the anomaly at $B^*$.}
\end{figure}

The most remarkable result is obtained using the AC technique with the field applied along the $c$ axis. For this orientation, we observed a weak but distinct anomaly at $B^*$, as shown in Fig.~\ref{fig:BalongcAC}(a). The exact position of the anomaly is defined from the second derivative of the specific heat with respect to the magnetic field, where the anomaly manifests itself as a small maximum, as shown in Fig.~\ref{fig:BalongcAC}(b). This anomaly was not observed in previous high-field specific heat measurements.

We will now discuss a possible origin of the high-field state above $B^*$ based on our findings. The presence of the specific heat anomaly at $B^*$ implies that it is likely a real thermodynamic phase transition rather than a crossover, contrary to what was previously suggested~\cite{Rosa2019}. The latter suggestion, however, was based on magnetostriction measurements performed in a magnetic field applied at 20$^\circ$ from the $c$ axis. Furthermore, the anomaly we observe at $B^*$ does not have the characteristic $\lambda$-like shape of a second-order phase transition contrary to those at $B_2$ and $B_N$ in Fig. \ref{fig:Balonga}(d). Therefore, the anomaly observed at $B^*$ most likely corresponds to a first-order phase transition. Moreover, it is thermodynamically forbidden that three second-order phase boundary lines meet at a triple point \cite{Yip1991}. This further supports our hypothesis that $B^*$ is a first-order phase transition. Finally, the anomaly at $B^*$ is observed only within the AFM state, in agreement with previous reports~\cite{Jiao2015,Moll2015,Ronning2017,Rosa2019,Kurihara2020}. Based on this, the most natural explanation of the phase transition at $B^*$ is a change of magnetic structure. Previous high-field NMR measurements unambiguously suggest that the AFM phases both below and above $B^*$ are incommensurate~\cite{Lesseux2020}. Therefore, $B^*$ should correspond to a transition from one incommensurate phase, AFM1, to another phase incommensurate along the $c$ axis, AFM4, with a propagation vector $\mathbf{Q} = (0.5, 0.5, l)$, where $l$ is different from 0.297 of the AFM1 phase.

This hypothesis is consistent with previous reports. Indeed, the previously observed resistivity jump at $B^*$~\cite{Moll2015,Ronning2017,Helm2020} can be naturally accounted for by a metamagnetic spin reorientation, as we suggest here. The only previously reported result, which is difficult to reconcile with our hypothesis, is that a Fermi surface reconstruction corresponding to the delocalization of the $f$ electrons occurs at $B^*$~\cite{Jiao2015,Jiao2017}. This conclusion, however, was challenged by recent angular-dependent de Haas-van Alphen effect measurements, which suggest that the $f$ electrons in CeRhIn$_5$ remain localized up to fields higher even than $B_c$~\cite{Mishra2021}.

\begin{figure}[htb]
\includegraphics[width=\columnwidth]{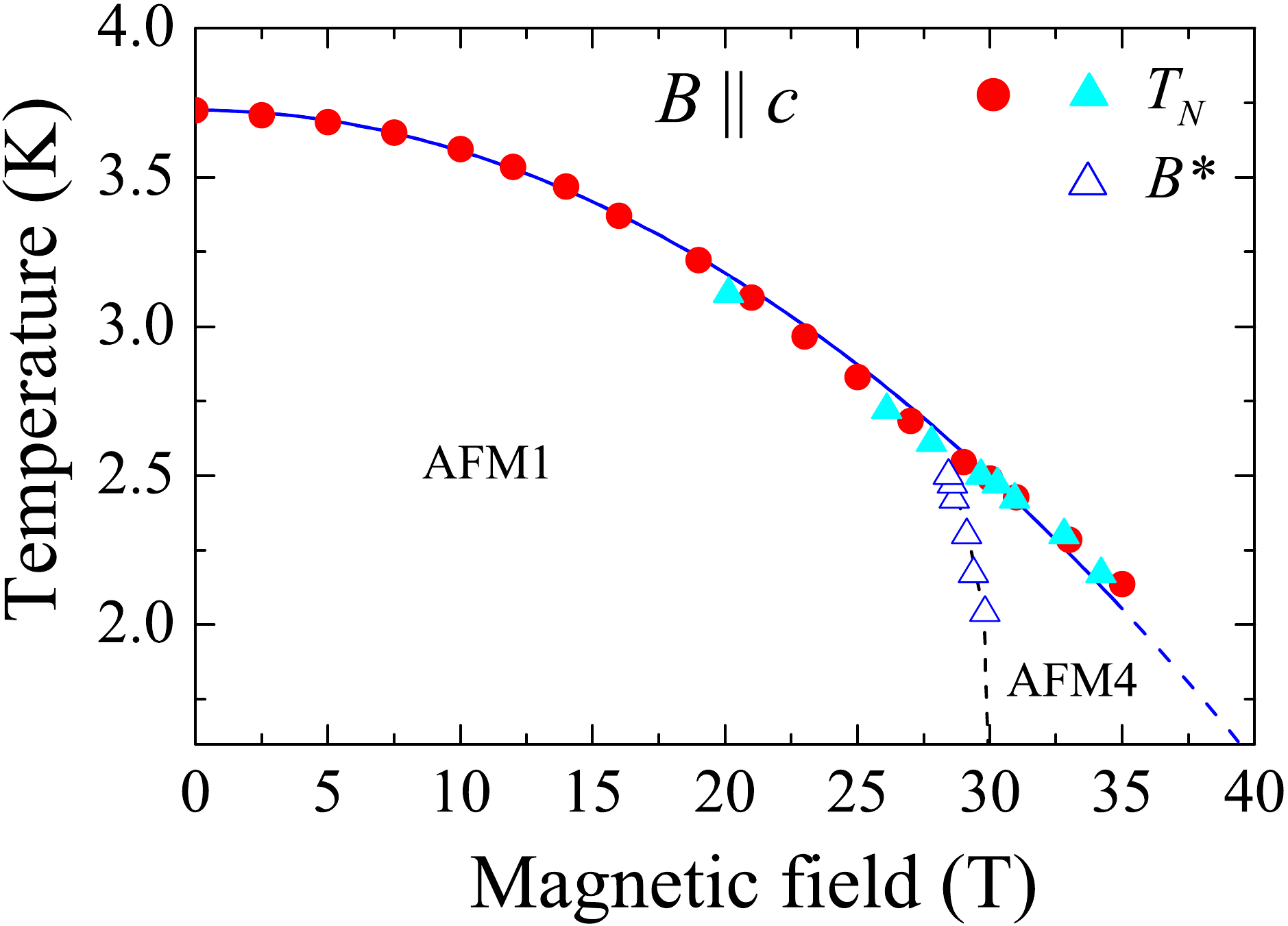}
\caption{\label{fig:PDc} Magnetic phase diagram of CeRhIn$_5$ obtained from relaxation (circles) and AC (triangles) specific heat measurements. Closed symbols correspond to second-order transitions from AFM to PM phase. Open symbols indicate the anomaly at $B^*$, which presumably corresponds to a weakly first-order transition.}
\end{figure}

Figure~\ref{fig:PDc} shows the revised magnetic phase diagram of CeRhIn$_5$ for a field applied along the $c$ axis. The field dependence of $T_N$ obtained from our static-field measurements is consistent with that previously reported, based on the pulsed-field data~\cite{Jiao2019}. A fit of the data to the $T_N(B) = T_{N0}[1 - (B/B_c)^2]$ expression, where $T_{N0}$ is $T_N$ at a zero field, reveals a critical field $B_c \simeq$ 52~T. This value is in agreement with that previously reported~\cite{Jiao2015}. The transition at $B^*$ is weakly temperature-dependent in agreement with previous measurements~\cite{Jiao2015,Moll2015,Ronning2017,Rosa2019,Kurihara2020}. As was already discussed above, we suggest that this first-order transition separates two different incommensurate magnetic phases.

We note that the situation is entirely different when a magnetic field is tilted from the $c$ axis. First, a finite component of a magnetic field in the basal plane explicitly breaks the $C_4$ rotational symmetry. Even more important is that at angles bigger than 2$^\circ$, the transition from incommensurate AFM1 phase into the commensurate phase AFM3 occurs below 30~T~\cite{Mishra2021}. Therefore, the transition at $B^*$ is from the commensurate phase AFM3 to the incommensurate phase AFM4. This is likely what was observed in recent ultrasound velocity measurements~\cite{Kurihara2020}. In these measurements, the anomaly observed at 20~T at the AFM1-AFM3 transition is very similar to that observed at $B^* \simeq$ 30~T. Furthermore, the magnetostriction anomaly observed at $B^*$ in a magnetic field tilted by about 20$^\circ$ from the $c$ axis is similar to that observed at 7.5~T, where it corresponds to the AFM1-AFM3 transition.

\section{CONCLUSIONS}

In summary, we performed specific heat measurements in CeRhIn$_5$ in static fields up to 36~T applied both along the $a$ and the $c$ axis. For the field along the $a$ axis, we confirmed the previously established rich phase diagram, and extended it to higher fields. For the field along the $c$ axis, we observed a distinct anomaly at $B^* \simeq$ 30~T, suggesting that a real thermodynamic phase transition, probably weakly first-order, is likely to take place at this field. We suggest that this transition is from the low-field incommensurate magnetic structure to another incommensurate phase, characterized by a different propagation vector. High field inelastic neutron scattering measurements are required to definitely confirm this hypothesis. Such measurements, although very challenging, are now possible due to the recent experimental breakthrough~\cite{Duc2018}.

\begin{acknowledgments}
We thank H.~Harima and Y.~Tokunaga for fruitful and enlightening discussions. We acknowledge the support of the LNCMI-CNRS, member of the European Magnetic Field Laboratory (EMFL), the ANR-DFG grant ``Fermi-NESt,'' and JSPS KAKENHI grants number JP15H05882, JP15H05884, JP15H05886, and JP15K21732 (J-Physics).
\end{acknowledgments}

\bibliography{CeRhIn5_specific_heat}

\begin{thebibliography}{28}
\bibitem [{Sup()}]{Suppl}%
  \BibitemOpen
  \href@noop {} {\bibinfo {title} {See {S}upplemental {M}aterial at [{URL}] for
  the experimental details, which includes {R}efs. {[22--24]}.}}\BibitemShut
  {Stop}%
\end{thebibliography}

\clearpage

\section*{Supplemental Material for ``Specific heat of CeRhIn$_5$ in high magnetic fields: Magnetic phase diagram revisited''}

\subsection{Specific heat method}

\begin{figure}[h]
  \includegraphics[width=\columnwidth]{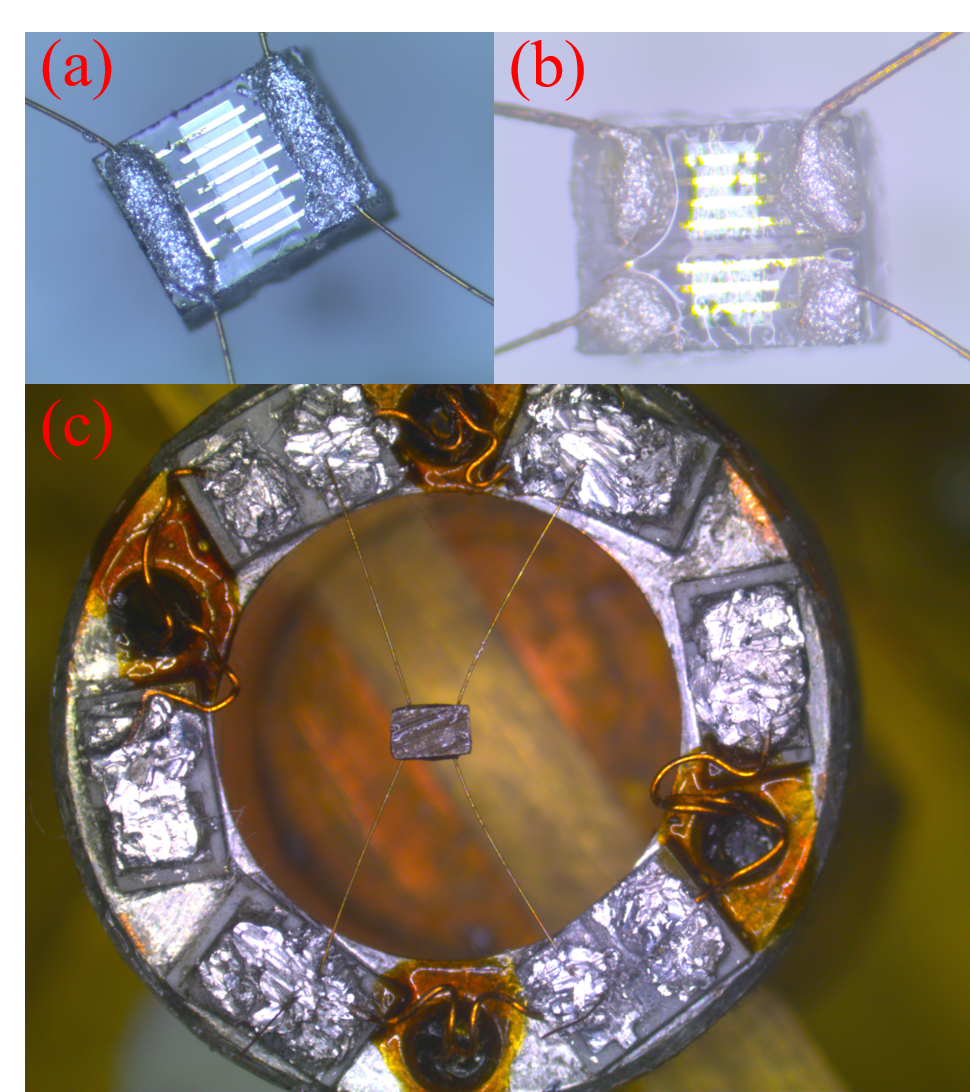}
  \caption{Setup used for specific heat measurement. (a) Cernox bare chip with four contacts used in the  thermal relaxation technique. (b) Cernox bare chip cut into two parts separating the heater and the thermometer for the AC calorimetric technique. (c) Single crystal of CeRhIn$_{5}$ mounted on the back of the Cernox bare chip connected to the thermal bath.}\label{fig:Cpsetup}
\end{figure}

Low-temperature specific heat measurements were performed in magnetic fields up to 36~T using both thermal relaxation and AC calorimetric techniques. Low-field measurements using thermal relaxation technique were performed in a $^{4}$He cryostat equipped with a 12~T superconducting magnetic and a VTI providing the lowest thermal bath temperature of 1.5~K.  The AC calorimetric measurements and the high field measurements using thermal relaxation technique were performed in a 36~T resistive magnet equipped with a $^4$He cryostat reaching the lowest thermal bath temperature of 1.3~K by pumping on the Helium bath. A Cernox thermometer, calibrated from 1.3~K to 40~K in magnetic fields up to 36~T, was used as a reference thermometer for the thermal bath.

\subsection{Thermal relaxation technique}

In the thermal relaxation method, a single Cernox bare chip, such as shown in Fig.~\ref{fig:Cpsetup}(a), was used as both the sample heater and the thermometer. The chip is connected to the thermal bath via a weak thermal link provided by four thin phosphor bronze wires, as shown in Fig.~\ref{fig:Cpsetup}(c). The wires also provide a mechanical support for the chip, and are used as the current leads. The sample is mounted on the back of the chip using a small amount of Apiezon grease, as shown in Fig.~\ref{fig:Cpsetup}(c). Further details of this technique are given elsewhere~\cite{Lortz2007}. Here we use long relaxations with a temperature increase of $100\%$ above the reference thermal bath temperature yielding a larger number of data points per relaxation. The resistance of the Cernox chip is calibrated in-situ against the thermal bath temperature using the reference thermometer. Similarly, the thermal conductance of the wires is calibrated in-situ against the thermal bath.  This technique provides an accuracy of 1$\%$, and its sensitivity of $10^{-3}$ is ideal to detect even very small changes in specific heat.

The Cernox chip, the Apiezon grease, and the wires contribute, as an addenda, to the total heat capacity of the system. This addenda must be subtracted to obtain the absolute specific heat of the sample. To this end, the addenda, i.e., the heat capacity of the system without sample, was first measured as a function of temperature in zero field. The addenda was later subtracted from all the curves obtained with a mounted sample.

\subsection{AC calorimetric technique}

Unlike the thermal relaxation technique, in AC calorimetry the heater and the thermometer are separated, as shown in Fig.~\ref{fig:Cpsetup}(b). The heater part of the Cernox chip of resistance $R_{H}$ is excited by a small AC current $I_{H}\cos(\omega t)$ of a few $\mu$A at a frequency $f =$ 4~Hz ($\omega = 2 \pi f$). The sample temperature is measured using the thermometer part of the chip excited by a small DC current of a few $\mu$A. The AC power generated by the heater has two effects. First, it increases the sample temperature above the thermal bath temperature $T_{bath}$. Second, it induces an oscillatory thermal response $T_{AC}$ with the sample temperature $T$ oscillating at twice the excitation frequency~\cite{Sullivan1968, Baloga1977}. The resulting sample temperature is thus given by:

\begin{equation}\label{sampleT}
T = T_{bath} + T_{DC} + T_{AC}
\end{equation}

Here $T_{DC}$ is an inevitable effect arising from self-heating of the heater and the thermometer due to supplied excitations. $T_{DC}$ is the excess in sample temperature $T$  with respect to the thermal bath temperature. It depends on the average power generated by the heater, $P_{H} = I_{H}^2R_{H}/2$, and the thermometer, $P_{Th}$, as well as the thermal conductance $\kappa$ between the system and the thermal bath, i.e., $T_{DC} = (P_{Th} + P_{H})/\kappa$.

The heat capacity of the system, $C$, is obtained from the oscillatory thermal response expressed in complex form as

\begin{equation}\label{deltaT}
T_{AC} = \frac{P_{H}}{\kappa + 2 \iota \omega C}
\end{equation}

Both the real and the imaginary terms of the complex Eq.~\ref{deltaT} depend on $P_{H}$, $\kappa$, $C$, $\omega$. However, the phase, $\theta$, between the two terms provides a more direct measure of the heat capacity of the system, independent of $P_{H}$:

\begin{equation}\label{tanq}
\tan\theta = \frac{2C\omega}{\kappa}
\end{equation}
In the first-order approximation, $\kappa$ is proportional to the temperature, and is independent of the magnetic field.

To precisely determine the sample temperature, the resistance of the Cernox heater is calibrated in magnetic field. To take into account the self heating effects, i.e., $T_{DC}$, the heater resistance at zero excitation $R_{H0}$ is obtained by a linear extrapolation of heater resistances $R_{H}$ at a few different small excitations. At zero self heating, the heater temperature is the same as bath temperature. Once $R_{H0}$ is known, the heater can be precisely calibrated in magnetic field using the field calibrated reference thermometer of the thermal bath. We determined $R_{H0}$ values at four different magnetic fields (20~T, 25~T, 30~T and 35~T) at several different temperatures in the range of interest of this study, i.e., from 1.3~K to 4~K. This way, calibration curves, i.e., $R_{H0}$ vs $T$, are obtained at above mentioned magnetic fields. Finally, using these calibration curves with appropriate polynomial fits, the sample temperature in magnetic field is precisely determined. Due to the magnetoresistance of the heater and the thermometer, there is a very small variation, less than 1\%, of the sample temperature in magnetic fields. The sample temperature, $T$, indicated in figures 1(d) and 3(b) of the main text is the value averaged over the relevant field interval.

\subsection{Comparison of thermal-relaxation and AC calorimetric techniques}

\begin{figure}[h]
  \centering
  \includegraphics[width=\columnwidth]{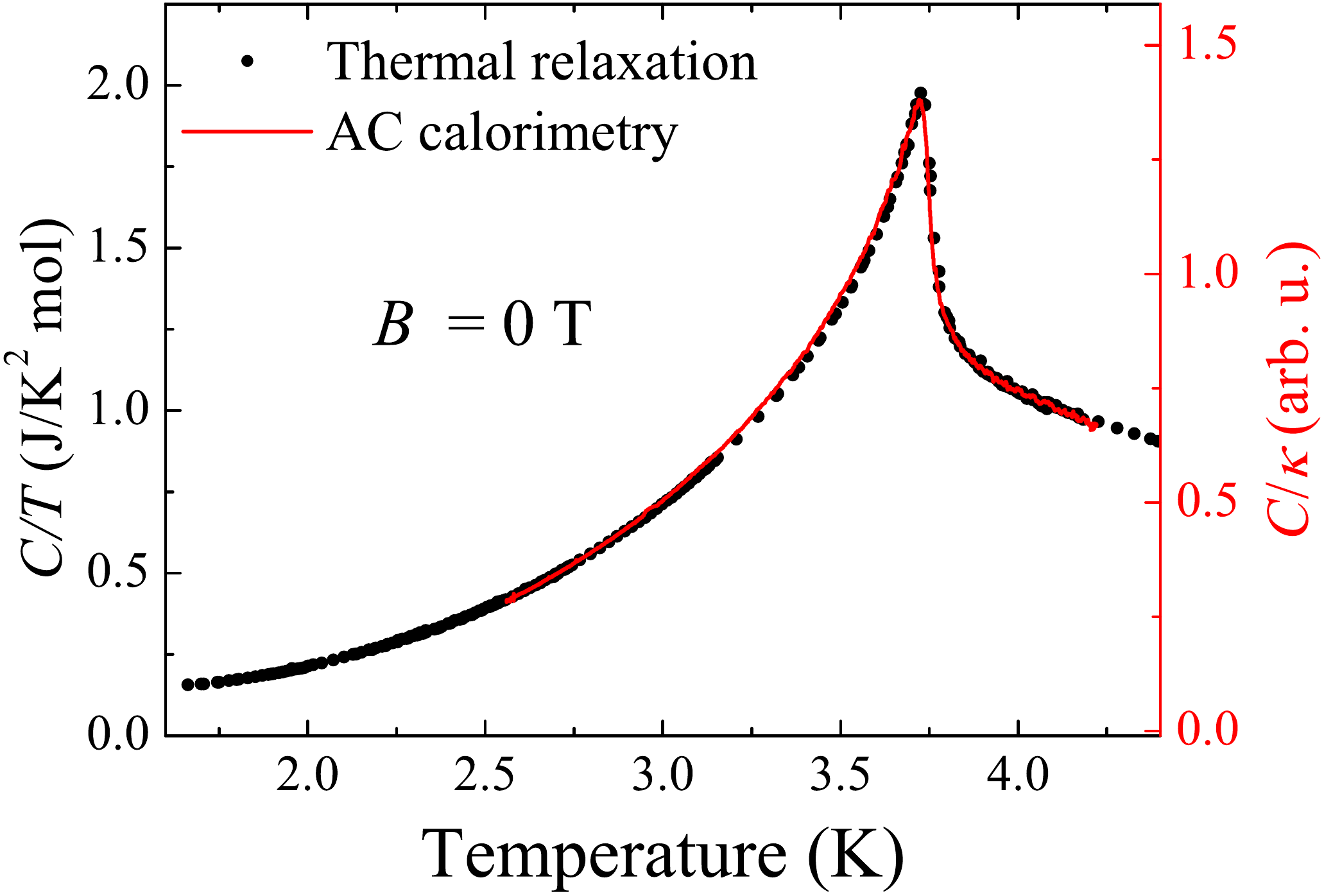}
  \caption{Specific heat of CeRhIn$_{5}$ at zero field obtained using thermal relaxation (closed circles) and AC calorimetric (solid line) techniques. For the latter, $\tan\theta$ proportional to $C/T$ is plotted.}\label{fig:0Tcomparison}
\end{figure}

The specific heat measured using the AC calorimetric technique is consistent with that obtained by relaxation technique, as shown in Fig.~\ref{fig:0Tcomparison}, where we plot $\tan\theta$. Given that $\kappa \propto T$, $\tan\theta \propto C/T$. For the AC specific heat data, the sample temperature is determined by removing the self heating effect, i.e., $T_{DC}$, using the heater calibration described in the previous section.

In the thermal relaxation technique, $C$ is the molar specific heat of CeRhIn$_{5}$ obtained by subtracting the addenda contribution from the total heat capacity of the system. On the other hand, in AC calorimetric technique, $C$ is the total heat capacity of the system including the addenda. However, the addenda contribution to the heat capacity of the system is negligibly small over the temperatures range of Fig.~\ref{fig:0Tcomparison}, as was independently verified using the thermal-relaxation technique. That is why there is an excellent agreement between the curves obtained by thermal relaxation and AC calorimetric techniques, as shown in Fig.~\ref{fig:0Tcomparison}.

\end{document}